\documentclass[%
reprint,
superscriptaddress,
groupedaddress,
 amsmath,amssymb,
 aps,
pra,
]{revtex4-1}

\usepackage{graphicx}% Include figure files
\usepackage{dcolumn}% Align table columns on decimal point
\usepackage{bm}% bold math
\begin{document}

%\preprint{APS/123-QED}

\title{Fast and High-Fidelity Readout of Single Trapped-Ion Qubit via Machine Learning Methods}% Force line breaks with \\

\author{Zi-Han Ding}
 \homepage{Program available on GitHub: https://github.com/quantumira-\\cle/On\_board\_FNN\_qubit\_discrimination}
\author{Jin-Ming Cui}%
\email{jmcui@ustc.edu.cn}
\author{Yun-Feng Huang}
\author{Chuan-Feng Li}
\email{cfli@ustc.edu.cn}
\author{Tao Tu}
\author{Guang-Can Guo}
\affiliation{%
CAS Key Laboratory of Quantum Information, University of Science and Technology of China, Hefei 230026, China.
}%
\affiliation{%
CAS Center For Excellence in Quantum Information and Quantum Physics, University of Science and Technology of China, Hefei 230026, People’s Republic of China.
}%

\date{\today}

\begin{abstract}
In this work, we introduce machine learning methods to implement readout of a single qubit on $^{171}\mathrm{Yb^{+}}$ trapped-ion system. Different machine learning methods including convolutional neural networks and fully-connected neural networks  are compared with traditional methods in the tests. The results show that machine learning methods have higher fidelity, more robust readout results in relatively short time.
To obtain a 99\% readout fidelity, neural networks only take half of the detection time needed by traditional threshold or maximum likelihood methods. Furthermore, we implement the machine learning algorithms on hardware-based field-programmable gate arrays and an ARM processor. An average readout fidelity of 99.5\% (with $10^5$ magnitude trials) within 171 $\mu$s is demonstrated on the embedded hardware system for $^{171}\mathrm{Yb^{+}}$ ion trap.

% \begin{description}
% Secondary publications and information retrieval purposes.
% \item[PACS numbers]
% May be entered using the \verb+\pacs{#1}+ command.
% You may use the \texttt{description} environment to structure your abstract;
% use the optional argument of the \verb+\item+ command to give the category of each item. 
% \end{description}
\end{abstract}

% \pacs{Valid PACS appear here}% PACS, the Physics and Astronomy
\maketitle

%\tableofcontents

\section{\label{sec:level1}Introduction}

Quantum computer \cite{nielsen2002quantum,cirac1995quantum} requires high-quality quantum logic gates, including high accuracy qubits operations and readout. Fast and high-fidelity readout \cite{PhysRevLett.113.220501} of quantum states \cite{kelly2015state} is essential for fault-tolerant quantum computation. The state~(bright/dark) of a trapped ion is derived through the qubit readout process. 

Traditional methods for qubit state discrimination like the threshold method, maximum likelihood \cite{langer2006high} or adaptive maximum likelihood \cite{PhysRevLett.100.200502} cannot guarantee high-accuracy readout within short detection time. The threshold method simply discriminates a qubit state with the sum of photon counting sequence, regardless of the inner patterns of data sequences. However, the state flip  caused by long-time laser interaction during state detection, which makes the bright state of $^{171}\mathrm{Yb^{+}}$ `jump' to the dark state, is hardly recognized by the threshold method. The maximum likelihood method has the ability to recognize state flips during the detection period, therefore guarantees higher discrimination accuracy. 
However, detection time consumption for the statistical process of the maximum likelihood method is large. 
Furthermore, the threshold and maximum likelihood methods grasp relatively few inner patterns or little information contained in photon counting sequences, which makes them hard to guarantee high accuracy within shorter detection time. 

With recently popular machine learning tools \cite{lecun2015deep}, we design methods to guarantee both low time consumption and high fidelity readout on ions trap system. Experiments have shown that machine learning methods like neural networks are good at capturing features and patterns in a sequence of data. We experimentally tested our proposed machine learning assisted single qubit readout methods on a trapped
$^{171}\mathrm{Yb^{+}}$ ion, and obtained faster and higher fidelity readout than traditional methods. 

Real-time state discrimination is important in applications like fast feedback control of quantum state operations. Considering that the field-programmable gate arrays (FPGA) \cite{brown2012field} based system has the ability to process data faster \cite{yang2017alpha}, we implement real time state discrimination with machine learning algorithms on an embedded hardware system instead of traditional CPU/GPUs implemention \cite{misra2010artificial}. We apply FPGA-based preprocessing and ARM-based feedforward neural networks for fast qubit readout.
The time property and fidelity are experimentally measured on our embedded hardware qubit readout system. 
In recent researches, machine learning method has also been applied in multi-ion states readout \cite{seif2018machine}, to reduce the crosstalk error between different qubits. Algorithms including support vector machine are applied in classification of quantum measurement trajectories for improving qubit measurements in superconducting systems~\cite{PhysRevLett.114.200501}. As the machine learning algorithms are compatible with our current work, fast and high fidelity multi-qubit readout can be realized in the future. 

\begin{figure}[htbp]
	\centering
	\includegraphics[width=0.5\textwidth]{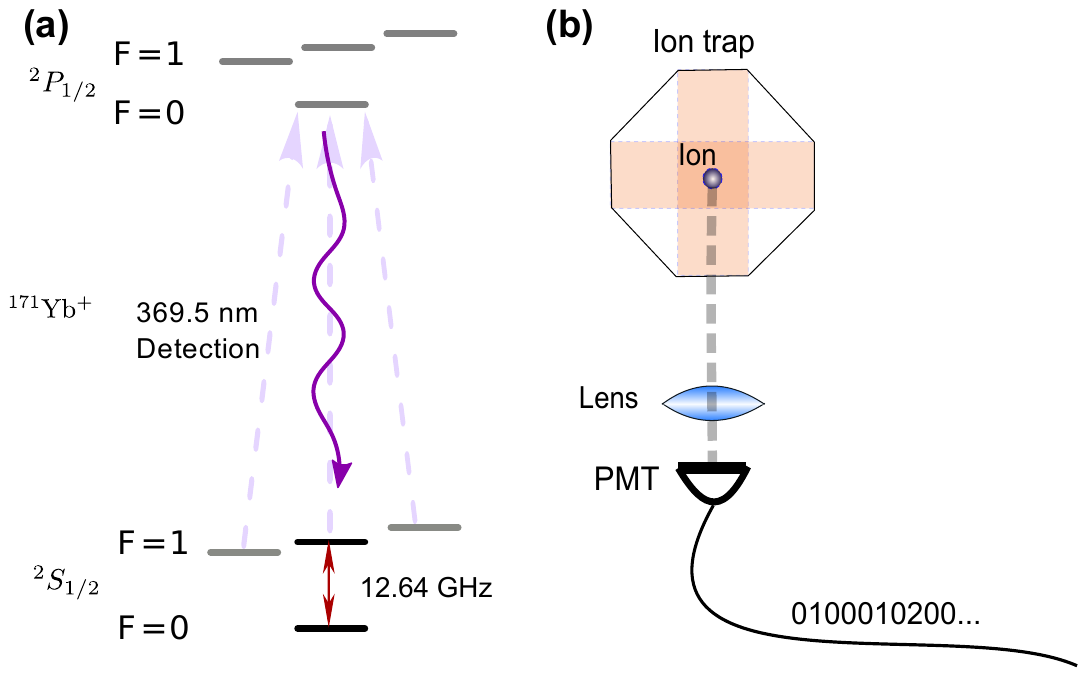}
	\caption{Energy levels and single qubit readout on $^{171}\mathrm{Yb^{+}}$ trapped-ion system. A laser of 369.53 nm is used to excite the bright state $^2S_{1/2}\left|F=1\right\rangle$, so as to spontaneously radiate fluorescence, which could be collected via a NA=0.4 lens and detected by a following photomultiplier tube (PMT). The output of PMT is processed to be photon counting numbers for the purpose of ion state (also called qubit) discrimination.}
	\label{fig:setup}
\end{figure}

\begin{figure*}[!tbp]
	\centering
	\includegraphics[width=1\textwidth]{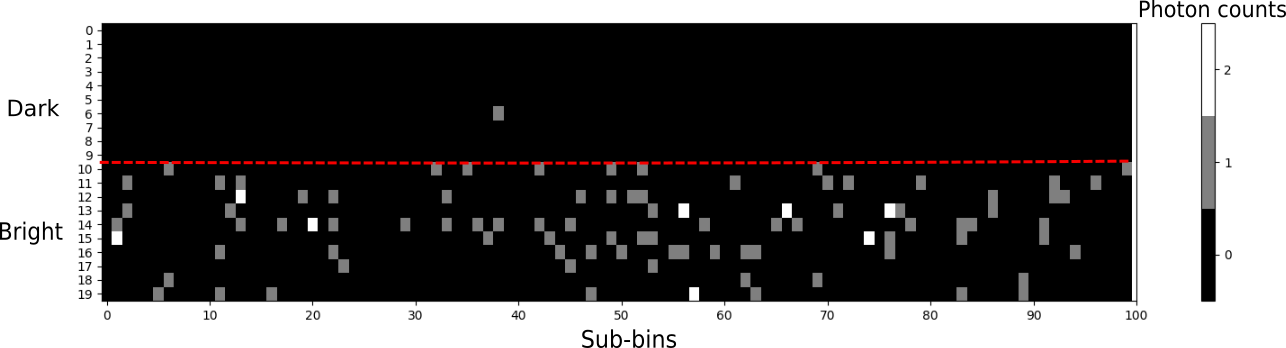}
	\caption{Single qubit dark states readout and bright states readout on the $^{171}\mathrm{Yb^{+}}$ trapped-ion system.
		The diagram shows 10-time readout experiment for both dart state (0-9) and bright state (10-19) sequences, each with photon counts in 100 sub-bins. }
	\label{fig:dark_bright}
\end{figure*}

\section{Machine learning method for qubit readout in ion trap}
The energy levels we use in $^{171}\mathrm{Yb^{+}}$ radiofrequency Paul trap \cite{debnath2016demonstration} are $^2S_{1/2}\left|F=0\right\rangle$ as the dark state and $^2S_{1/2}\left|F=1\right\rangle$ as the bright state, as shown in Fig.\,\ref{fig:setup}(a). The magnetic dipole transition between these two energy levels is corresponding to the 12.6 GHz microwave \cite{ospelkaus2011microwave} operation. The 369.53 nm detection laser can arouse resonance between two states: $^2S_{1/2}\left|F=1\right\rangle \leftrightarrow  ^2P_{1/2}\left|F=0\right\rangle$. A state flip could be caused by a microwave $\pi /2$-pulse. The excited state $^2P_{1/2}\left|F=0\right\rangle$ will keep a spontaneous radiation process to emit fluorescence, which could be detected by the photomultiplier tubes (PMT). However, the state $^2S_{1/2}\left|F=0\right\rangle$ cannot be excited because of the 12.6 GHz detuning. The readout process of single qubit is shown in Fig.\,\ref{fig:setup}(b). We experimentally use an objective lens with NA=0.4 for detection of photons.

Visualized results of single qubit readout are shown in Fig.\,\ref{fig:dark_bright}. We experimentally measure dark state and bright state 10 times each with PMT, to get 100 sub-bins (3 $\mu$s per sub-bin) data sequence each time. These sequences are raw data used to determine whether it is a dark or bright state qubit. 

Machine learning methods are proved to have prominent effects on applications like sequence data classification. We test several machine learning methods including fully-connected neural network (NN), convolutional neural network (CNN) \cite{krizhevsky2012imagenet}, recurrent neural network (RNN) \cite{hochreiter1997long}, support vector machine (SVM) \cite{hearst1998support}, logistic regression \cite{hosmer2013applied}, K-neighbors classifier \cite{keller1985fuzzy} and decision tree classifier \cite{safavian1991survey} to compare with traditional methods like threshold method and maximum likelihood method \cite{PhysRevLett.100.200502}.

\begin{figure}[bp]
	\centering
	\includegraphics[width=0.5\textwidth]{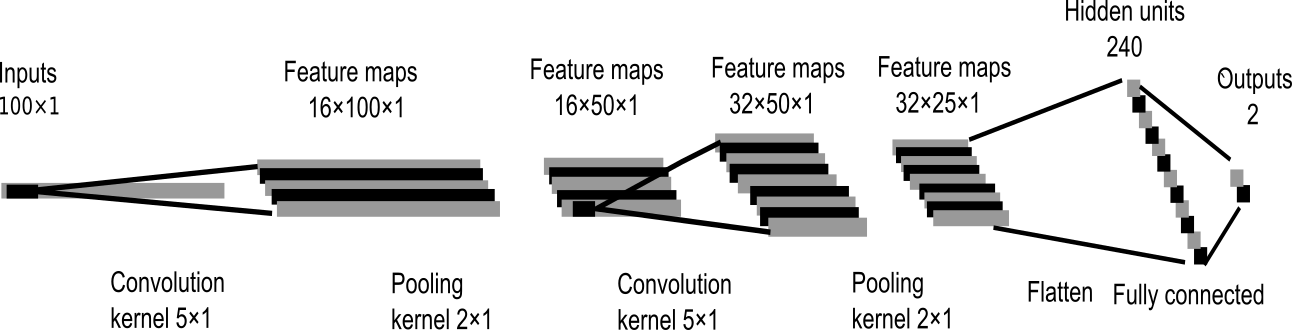}
	\caption{Architecture of CNN for single qubit readout in test. The 1D CNN has two convolutional layers of kernel $5\times 1$, two max-pooling layers of size $2\times1$, and two fully-connected layers of 240 and 2 units respectively. For  the two units of output layer: $y_1$ and $y_2$, if $y_1>y_2$,  the state  is inferred as bright state; otherwise, it is a dark state.}
	\label{fig:cnn_network}
\end{figure}

Fig.\,\ref{fig:cnn_network} shows the architecture of 1D CNN we applied for experiments, similar with LeNet-5 \cite{lecun1998gradient}. It has two convolutional layers, two pooling layers and two fully-connected layers. The dimension of input is 100$\times$1 (downsampling into 50$\times$1, 25$\times$1 through max-pooling of size 2$\times$1 after each convolution). The first convolutional layer has 16 feature maps of size 100$\times$1, with convolution kernel 5$\times$1.  The second convolutional layer has 32 feature maps of size 50$\times$1. Neural units in the fully-connected layer and the output layer are of size 240 and size 2, respectively. For gradient descent process, the AdamOptimizer \cite{kingma2014adam} is applied with the decayed learning rate of $1\times10^{-3}\sim 1\times10^{-4}$, so that the model could be trained to reach optimum within $2\times10^5$ samples.  The loss function we apply in our model is the absolute difference between output inferences and target labels (with similar learning effect as cross entropy loss in our model) as Eq. (1) 
\begin{equation}
 Loss=\sum_{x\in X}abs(y_x-\hat{y_x})	
\end{equation}
where $X$ is the set of input photon counts sequences, $y_x$ is the output inference values of neural network and $\hat{y_x}$ is the true label of the corresponding input sequence.

Structures of the other machine learning methods are not further described here, but they are all experimentally optimized. Inputs of machine learning methods are original photon counting sequences, which are 100 sub-bins counts in detection for each state. Training data is generated directly on the experiment systems. With fast preparations of bright and dark states on ions trap, we can obtain a large batch of training data in a short period of time. After training the CNN model with experimental data, we use it to classify the states generated in experiments afterward. The outputs of CNN are labels of bright or dark states. Therefore, CNN is an end-to-end model for qubit discrimination, as well as the other machine learning methods.

We experimentally tested our proposed machine learning methods and traditional methods. As shown in TABLE \uppercase\expandafter{\romannumeral1}, among all machine learning and traditional methods, CNN method achieves the highest accuracy rate and relatively robust performance with the input of 100 sub-bins photon counts data. 
\begin{table}[bp]
	\caption{\label{tab:diff_compare}Comparison of accuracy and inference time of different single qubit discrimination methods}
	\begin{ruledtabular}
		\begin{tabular}{ccddd}
			Methods&Accuracy(\%)&\mbox{\qquad Time(seconds)}\\
			\hline
			Threshold&99.248$\pm0.07$&\mbox{1.563}\\
			Maximum Likelihood&99.311$\pm0.12$&\mbox{4678.839}\\
			Fully-connected NN
			&99.411$\pm0.13$&\mbox{{2.865}} \\
			$\mathbf{CNN}$&$\mathbf{99.413\pm0.10}$&\mbox{7.065}\\
			RNN&99.364$\pm0.10$&\mbox{914.113}\\
			SVM&99.341$\pm0.10$&\mbox{4.282}\\
			Logistic Regression&99.123$\pm0.11$&\mbox{4.137}\\
			K-neighbors Classifier&92.125$\pm0.14$&\mbox{3965.299}\\
			Decision Tree Classifier&98.010$\pm0.30$&\mbox{1.504}\\
		\end{tabular}
	\end{ruledtabular}
\end{table}
To investigate the time property of different methods, a test with a dataset of $2\times10^5$ samples were undertaken for each method. The time consumption of CPU computation for qubit discrimination is also shown in TABLE \uppercase\expandafter{\romannumeral1}. The faster the inference process needs, the sooner we can get the states of the ion. Traditional methods like maximum likelihood need long computation time for complete numeric statistics of photon bin counts to keep high accuracy.  However, most ML methods demonstrate a lower demand on computation time, except for RNN and K-neighbors Classifier methods. With a significantly higher accuracy of qubit discrimination, CNN and fully-connected neural network methods only need relatively short computation time, which makes them promising for fast readout and feedback control of single qubit. We will further speed up the neural network method with hardware implementation in Section \uppercase\expandafter{\romannumeral3}.

\begin{figure}[bp]
	\centering
	\includegraphics[width=0.5\textwidth]{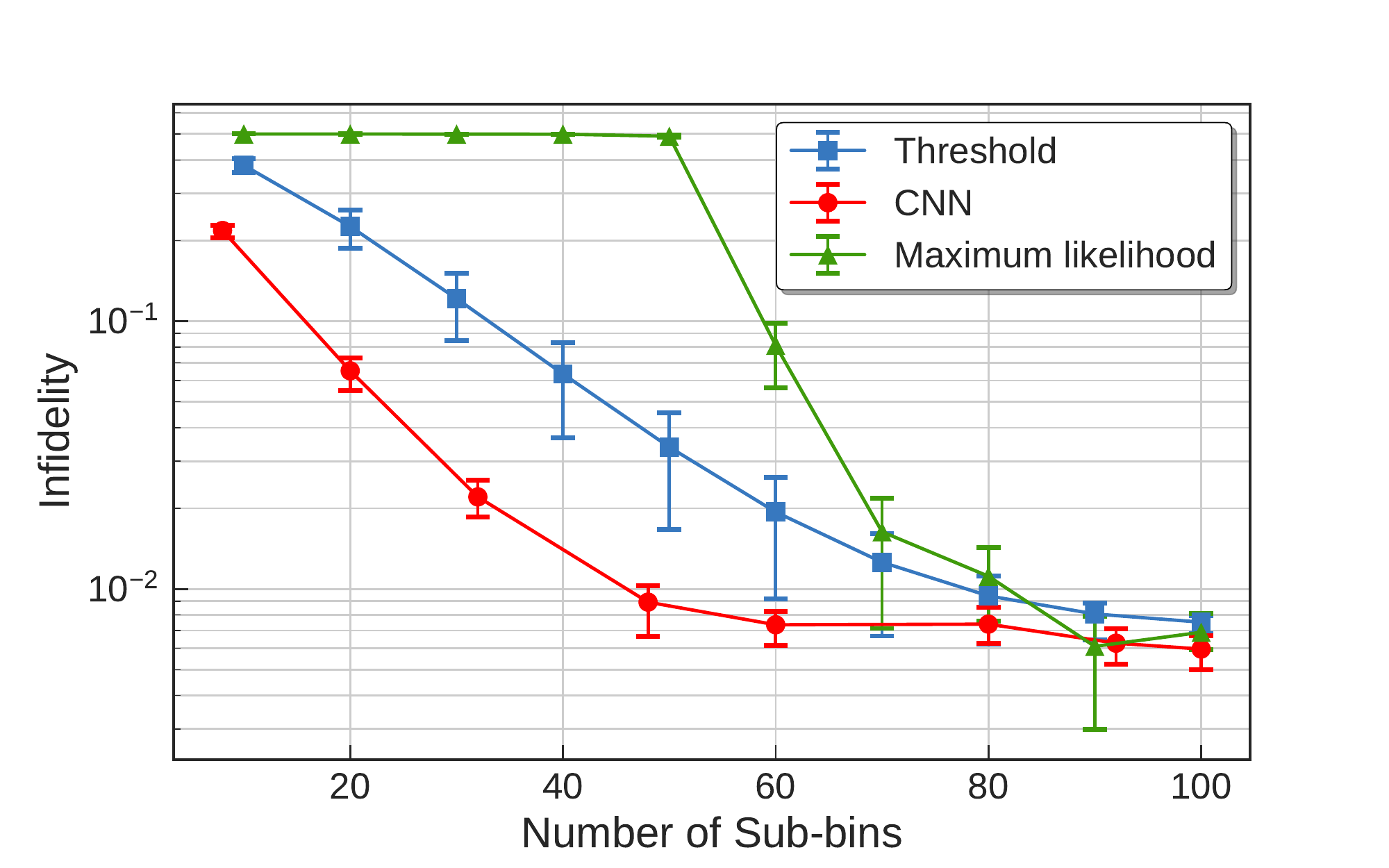}
	\caption{Infidelity of single qubit readout with different number of sub-bins in test. The proposed CNN method can achieve 99\% accuracy within 43 sub-bins ($\sim$ 129 $\mu$s), while threshold and maximum likelihood methods need more than 80 sub-bins. And CNN has more robust readout performance. }
	\label{fig:cmp_diff_methods}
\end{figure}

Fig.\,\ref{fig:cmp_diff_methods} shows the relationship between infidelity (=1-fidelity) and the number of sub-bins (fixed sub-bin time) for CNN and two traditional methods. To achieve 99\% accuracy, only 43 sub-bins ($\sim$ 129 $\mu$s) are needed by CNN method, while the threshold method needs 80 sub-bins ($\sim$ 240 $\mu$s). The proposed CNN method only needs half the amount of data needed by traditional methods to guarantee the same accuracy, which significantly reduces the detection time needed by qubit discrimination. With fewer bins data, ML methods could make qubit discrimination process even faster. Additionally, the error bars in the diagram confirms the robust performance of CNN compared with the other two methods, which is essential for high-fidelity one-shot single qubit readout.

\section{\label{sec:level1}Embedded qubit readout system}
%above: simulation test on PC; below: experiment on board
To discriminate the state in real time with single shot readout, we need to record the photon counts and conduct the feedforward neural network algorithm in each detection. Special hardware is needed to support this feature. The overall computation process is embedded into the chip Zynq-AX7020, which has an ARM (Cortex-A9, 767 MHz) processor and FPGA (XC7Z020-2CLG400I). The Transistor-Transistor Logic (TTL) signals from PMT are directly sent to the chip. Two main functions are accomplished on chip: (1). signal transformation from TTL to digital photon counts and its storage on registers; (2). the feedforward neural networks implementation on ARM with FPGA computation acceleration. With the above settings, the average processing time for single sample discrimination is significantly decreased. 
\begin{figure}[htbp]
	\centering
	\includegraphics[width=0.5\textwidth]{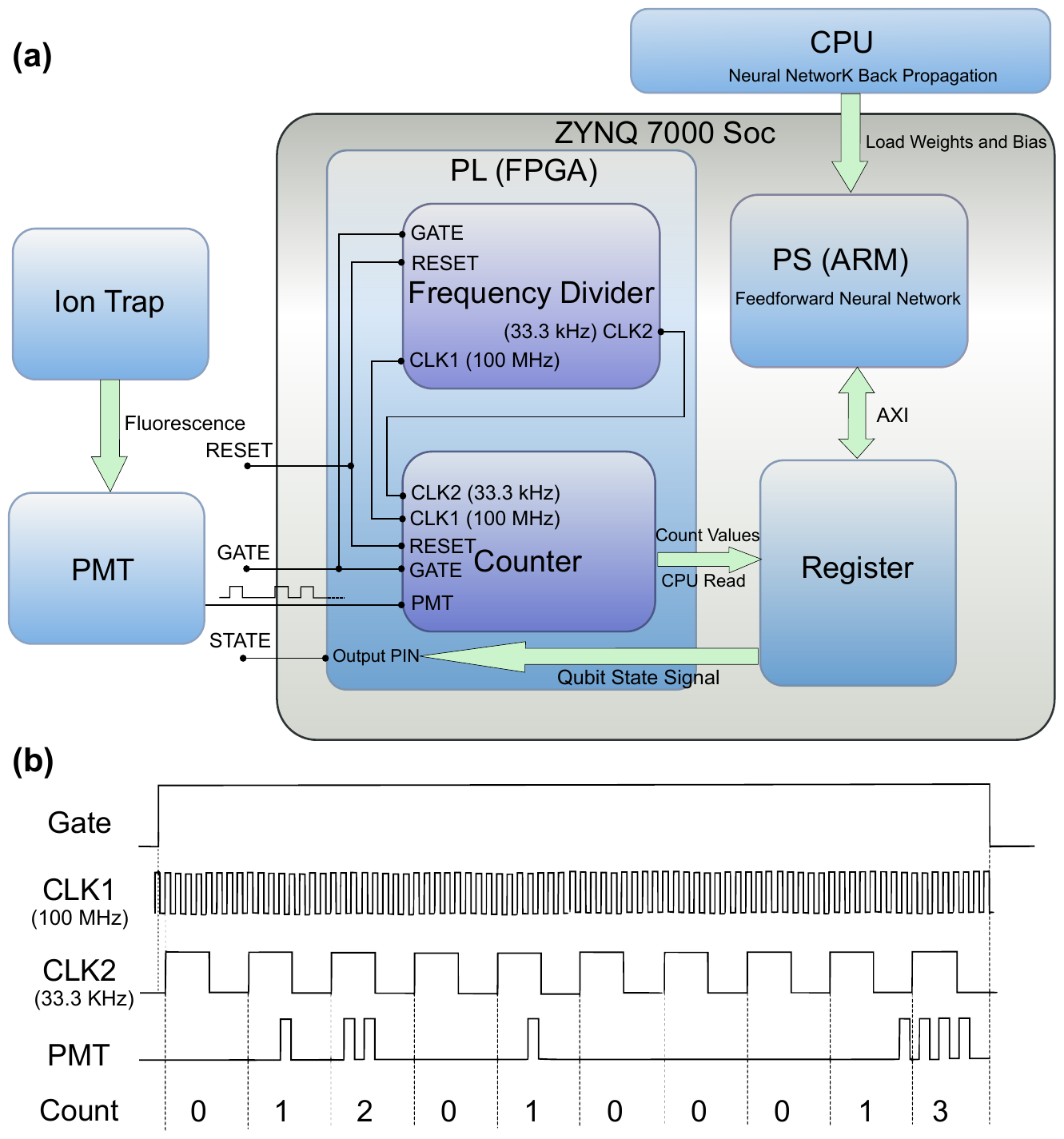}
	\caption{The diagram of fast qubit readout system in experiment. (a) Scheme of overall system. The inputs of our proposed embedded qubit readout system are TTL signals from the PMT, which is acquired through detecting the bright state fluorescence in the ion trap system. The gate signal is an enable/disable signal to control the counting process. A frequency divider and a counter are implemented on FPGA and the implementation of a feedforward neural network is programmed into ARM. Every time the counter counts a photon number in a sub-bin time, PL triggers an interrupt to stop the loop of software program in PS and sends the number through registers to PS. Parameters of weights and bias of the neural network are pre-trained on a CPU or GPU and loaded on Zynq development board. Whenever PS receives a signal of the end of a bin time, it automatically starts the inference process  and outputs the state of the ion to the pin. (b) Waveform sequence diagram of controlling signals in experiment. The 33.3 kHz clock is generated from the 100 MHz system clock, serving as the controlling signals of the counter. The extraneous 1.67 kHz gate signal sets the time window (300 $\mu$s high-level voltage for detecting time) of single-sample photons counting, and it is the controlling signals of both the counter and the frequency divider. Each rising edge of PMT signals is detected to be one photon count within one-sub-bin time 30 $\mu$s, which is in accordance with the period of generated 33.3 kHz clock.}
	\label{fig:FPGA_diagram}
\end{figure}

The diagram of the embedded system on Zynq-7000 development board for fast qubit readout is shown in Fig.\,\ref{fig:FPGA_diagram}(a). A frequency divider and an analog-to-digital counter for TTL signals are implemented on FPGA through hardware programming. Experimentally, we set the sub-bin time to be 30 $\mu$s instead of 3 $\mu$s used in the simulation tests above, to achieve faster qubit discrimination. Correspondingly, the number of input units becomes 10 instead of 100 to keep the total bin time unchanged. Our tests show that, to some extent, merging of inputs does not significantly decrease the effects of neural networks for qubit discrimination.  The system clock for the programmable logic unit(PL) is 100 MHz ($\sim$ 10 ns). The frequency divider transforms this clock to be around 33.3 kHz clock, which is used as a sampling clock to record  photon counts within sub-bin time 30 $\mu$s. As shown in Fig.\,\ref{fig:FPGA_diagram}(b), the 33.3 kHz clock is generated from the 100 MHz system clock on board through the frequency divider, starting from the positive edge of the gate controlling signal and ending with the negative edge of the gate signal. The maximum asynchronous error of the generated clock and the gate signal could be no more than 10 ns, which is the period of 100 MHz system clock. The generated clock, together with the gate signal, controls the time bin of counter. The falling edge of the gate signal will trigger an interrupt for the ARM to start the inference process and output the state of the ion.

The feedforward implementation of the fully-connected neural network with two layers of size 20 and size 2, respectively, and with rectified linear unit (ReLU) \cite{nair2010rectified} as activation function, is set on the processing system core unit (PS). $ReLU(x)=x$ when input $x >0$, otherwise $ReLU(x)=0$. The weights and bias parameters are pre-trained through the backpropagation \cite{lecun1988theoretical} process on CPU with a neural network exactly the same as the feedforward implementation on the PS. The AdamOptimizer is used for training this fully-connected neural network with the decayed learning rate of $1\times10^{-3}\sim 1\times10^{-4}$. After it is trained to be optimal with about $2\times 10^5$ samples, the weights and bias parameters are stored into files, to be loaded on the development board for inference process.

\section{Experimental results}

\begin{figure}[htbp]
	\centering
	\includegraphics[width=0.5\textwidth]{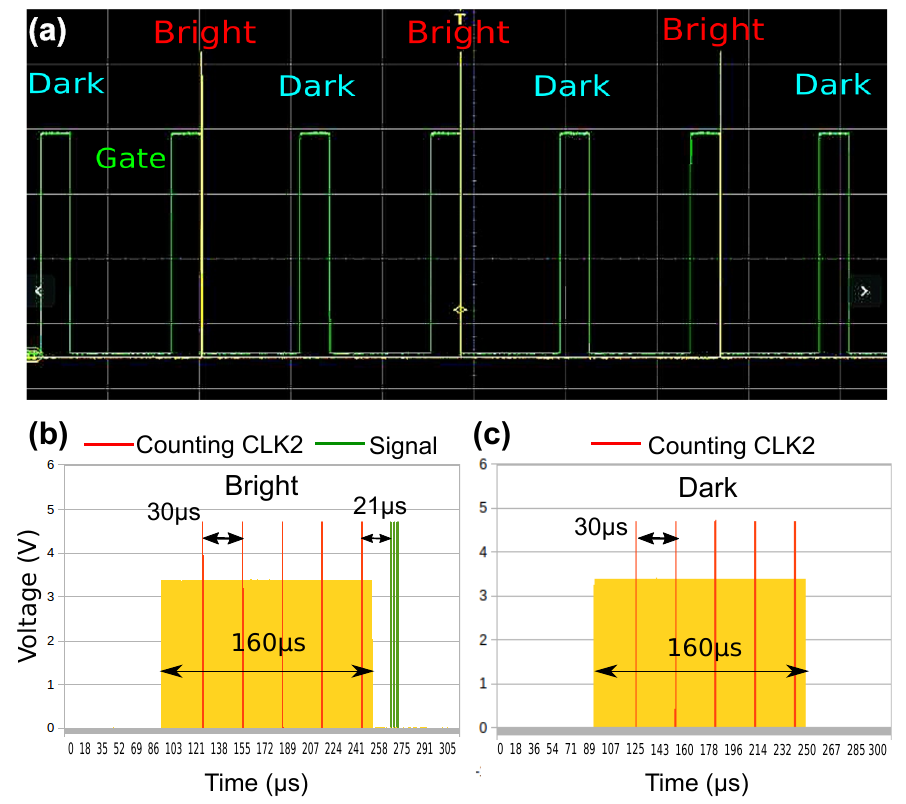}
	\caption{ Detection signals measured by an oscilloscope. (a) Experimentally readout of a crossfade bright/dark states sequence, with 7 states in scope and 5 detection sub-bins for each state. (b), (c) Zoom in signals of a dark state and a bright state, respectively. The yellow rectangular wave is the high-level gate signals of 160 $\mu$s ($>30\times5$, to make sure of 5 detection bins), the red 1 $\mu$s pulse inside gate signal is the sampling clock sign for each sub-bin. The green triple pulse right after the high-level gate signal in (b) is the signal of bright state, without which it indicates a dark state.}
	\label{fig:Experiment_oscilloscope}
\end{figure}
We experimentally test the methods above, using an oscilloscope to display the detection results of qubit readout in Fig.\,\ref{fig:Experiment_oscilloscope}. A bright-state/dark-state crossfade sequence is prepared with microwave $\pi /2$-pulse in the single trapped ion system, as shown in Fig.\,\ref{fig:Experiment_oscilloscope} (a). The detection of bright state is set to output a spike pulse signal after 5 sub-bin photons counting and 21 $\mu$s inference time. The signal is actually a triple 1 $\mu$s pulse following the high-level gate signals, seen more clearly in Fig.\,\ref{fig:Experiment_oscilloscope}(b). The zoom in readout result of single sample qubit readout for bright state and dark state is shown in Fig.\,\ref{fig:Experiment_oscilloscope}(b,~c), with extra signals of each counting period (five red pulses). We could derive that the inference time for each sample is around 21 $\mu$s on the embedded system, as measured in Fig.\,\ref{fig:Experiment_oscilloscope}(b). However, the average inference time with CPU/GPUs on PC is 72 $\mu$s in TABLE \uppercase\expandafter{\romannumeral2}, not to mention the considerable time consumption on communication between the PMT counter and the computer. Therefore, the embedded hardware implementation of fully-connected neural network method has a speed-up of over 3 times in comparison to general systems. The total time consumption for single sample qubit readout is about 171 $\mu$s (5 bins $\times$ 30 $\mu$s/bin + 21 $\mu$s = 171 $\mu$s).

\begin{figure}[htbp]
	\centering
	\includegraphics[width=0.5\textwidth]{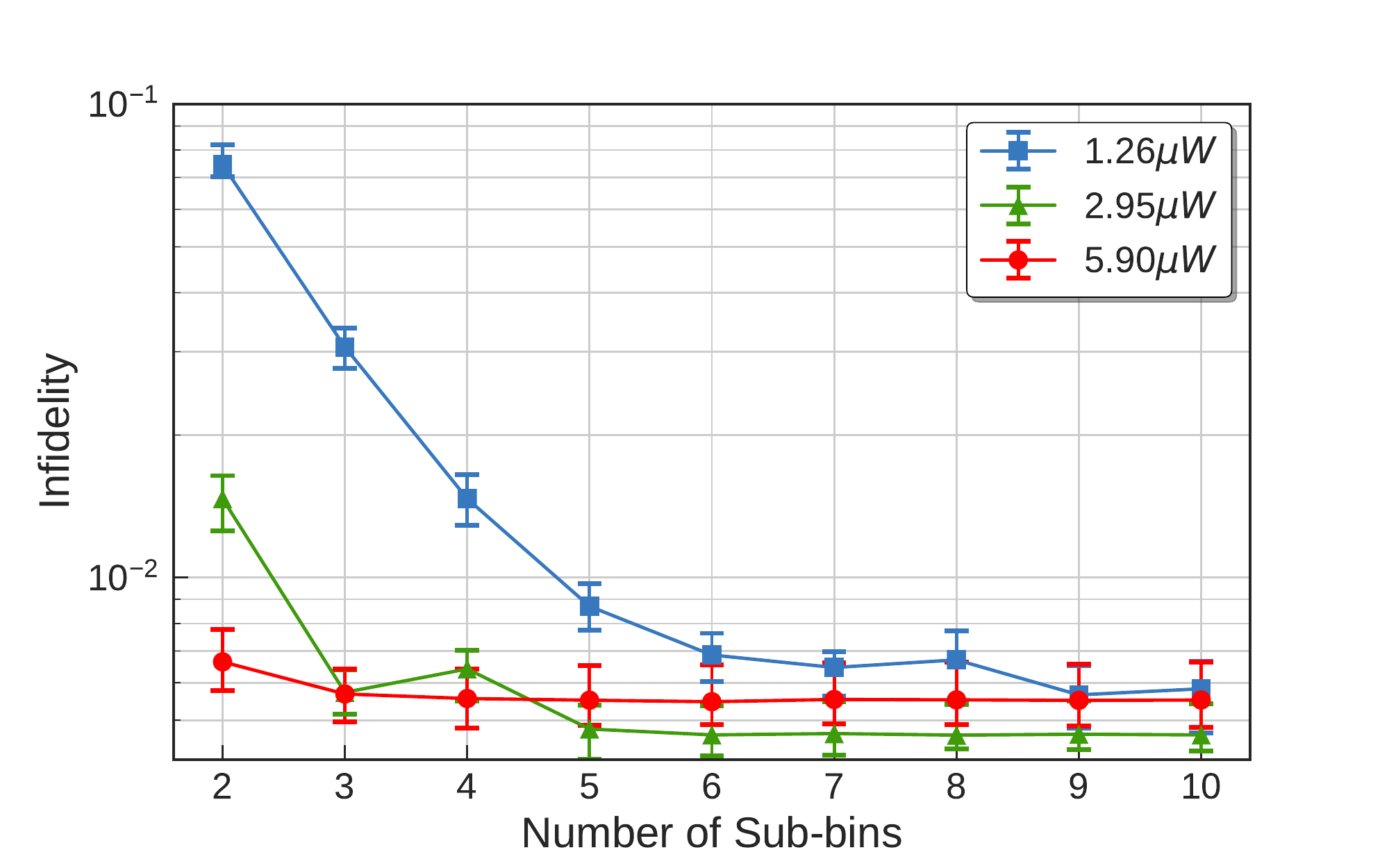}
	\caption{Infidelity of single qubit readout with different incident laser powers using fully-connected neural network method. Comparison of incident laser powers of 1.26 $\mu$W, 2.95 $\mu$W and 5.90 $\mu$W is shown. The laser power of 2.95 $\mu$W has the best performance of 99.53\% with only 5 bins ($\sim$ 150 $\mu$s detection time). }
	\label{fig:infidelity}
\end{figure}

\begin{figure}[bp]
	\centering
	\includegraphics[width=0.5\textwidth]{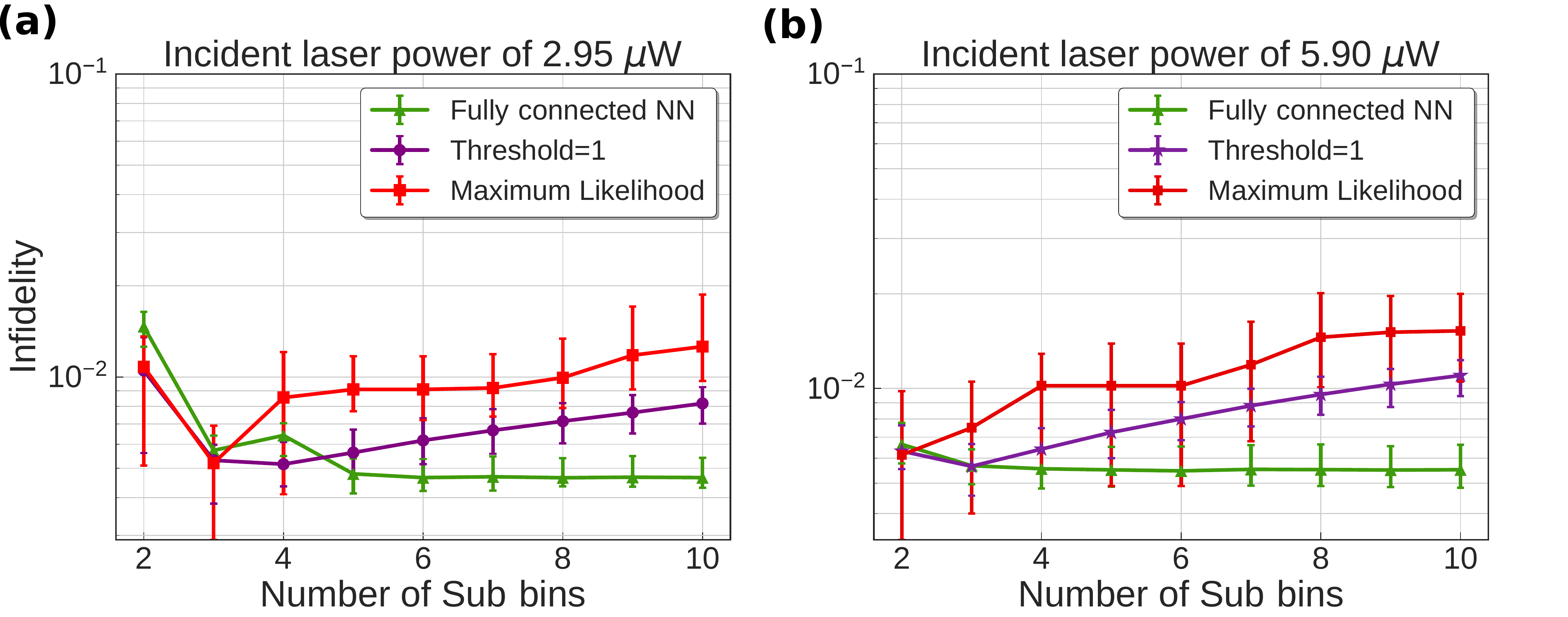}
	\caption{Comparison of fully-connected neural network method with threshold method and maximum likelihood method for single qubit readout with incident laser power of 2.95 $\mu$W and 5.90 $\mu$W. With the laser power of 2.95 $\mu$W, the fully-connected neural network method achieves the highest fidelity overall; with the laser power of 5.90 $\mu$W, the fully-connected neural network method shows higher fidelity and more robust performance than threshold method and maximum likelihood method. } 
	\label{fig:Compare_NN_Threhold}
\end{figure}

As there is a saturation  power for detecting the ion fluorescence, we might choose the optimum excitation laser power for detection. The detected fluorescence photon counting rate $n$ with excitation power $P$ is
\[
n(P) = n_0 \frac{x}{1+x} = n_0\frac{P/P_0}{1+P/P_0}, 
\]
where $P_0$ is saturation power, $n_0$ denotes saturation photon counting rate. In our setup, $P_0$ is measured as 2.91 $\mu$W, and $n_0$ is $1.39\times 10^5$ counts per second. 
The performance of fully-connected neural network with different numbers of sub-bins on hardware (fewer than 10) and different powers of incident 369.53 nm laser in trapped ions cavity is shown in Fig. \ref{fig:infidelity}. The best accuracy of 99.53$\%$ is achieved with the incident laser power of 2.95 $\mu$W. Moreover, a faster qubit readout with only 2 sub-bins data as input could be achieved using a stronger laser of 5.90 $\mu$W, as well as maintaining the fidelity higher than 99\%. Comparisons of the onboard fully-connected neural network method and the threshold method for single qubit readout are shown in Fig.\,\ref{fig:Compare_NN_Threhold}, with incident laser power of 2.95 $\mu$W and 5.90 $\mu$W respectively. The rise of infidelity for threshold method with more sub-bins shows that the dark state has larger chances to be excited into bright state with larger incident laser powers like 2.95 $\mu$W and 5.90 $\mu$W rather than using 1.26 $\mu$W as usual. More importantly, the robust performance of neural network method proves that it could capture these states transformation features, which is reflected by the photon counts sequence, in the process of state inference. Considering time property, fidelity and robustness, the best performance is achieved by the fully-connected neural networks on board with 5 sub-bins (30 $\mu$s for each bin) data as input variables and 2.95 $\mu$W 369.53 nm detecting laser in system.

\begin{table}[htbp]
	\caption{\label{tab:table2}Time property of different computation architectures for single sample qubit discrimination}
	\begin{ruledtabular}
		
		\begin{tabular}{c|p{2cm}<{\centering}p{2cm}<{\centering}}
			&\multicolumn{2}{c}{Computation Time(ms)} \\
			\hline
			Architectures&Loading files&\mbox{Inference}\\
			\hline
			CPU+GPU (Test)&0.42&\mbox{0.072}\\
			ARM (Test)&91.2&\mbox{2.7}\\
			ARM+FPGA (Test)&2.2&\mbox{1.1}\\
			ARM+FPGA (Experiment)& --- &\mbox{\bfseries0.021}\\
		\end{tabular}
	\end{ruledtabular}
\end{table}

The time property of different computation architectures for single qubit discrimination of one sample is shown in TABLE \uppercase\expandafter{\romannumeral2}. All architectures have realized the function of a feedforward fully-connected neural network with 10 input units, as described above. As shown in Fig.\,\ref{fig:FPGA_diagram}, the files containing weights and bias of the neural network are loaded into the computation modules, including GPU (test) and ARM (test/experiment). Then the computation of neural network inference is accomplished on these architectures. The three tests are implemented with data files of digital photon counting sequences, derived through measurements of ion state before the tests; the experiment with  ARM and FPGA on the Zynq-7000 development board is implemented with PL part for processing raw inputs of PMT signals from the physical system, which are analog instead of digital ones. In other words, ARM+FPGA in test only use the PS part of system in Fig.\,\ref{fig:FPGA_diagram}, while ARM+FPGA in experiment use both PL and PS parts. And FPGA in test is for acceleration of the PS computation process, different from in experiment.

\section{Conclusion}  
To conclude, for single qubit readout in trapped ions system, we can achieve 99.5\% fidelity within around 171 $\mu$s per sample, using FPGA and ARM-based fully-connected neural network method. Moreover, a higher fidelity and more robust performance could be achieved with convolutional neural networks and more sub-bins. Generally, the machine learning methods are applied to grasp more detailed features of photon counting sequences for a higher discrimination accuracy with shorter detection time; while the FPGA and ARM processor based hardware provides a faster process of neural networks inference. Furthermore, our embedded qubit readout method could be extended to  synchronous multi-ions states readout and fast feedback control of qubit states in trapped-ion systems, which is promising for more flexible quantum gates operations in the future.

\acknowledgments
We would like to thank Yao Teng, Chang-Qing Feng and Yan-Wen Wu for their experimental support and helpful discussions with us. This work is supported by the National Key Research and Development Program of China (No. 2017YFA0304100, 2016YFA0302700), the National Natural Science Foundation of China (Nos. 61327901, 11774335, 11474268, 11474270, 11734015, 11821404), Key Research Program of Frontier Sciences, CAS (No. QYZDY-SSW-SLH003), the Fundamental Research Funds for the Central Universities (No. WK2470000026, WK2470000018), Anhui Initiative in Quantum Information Technologies (AHY020100, AHY070000), the National Program for Support of Topnotch Young Professionals (Grant No. BB2470000005).

\bibliography{ref}

\end{document}